\documentclass[showpacs,preprintnumbers,amsmath,amssymb]{revtex4}
\usepackage{bm}

\newcommand{\be}{\begin{equation}}
\newcommand{\ee}{\end{equation}}
\newcommand{\bea}{\begin{eqnarray}}
\newcommand{\eea}{\end{eqnarray}}
\newcommand{\nn}{\nonumber}
\newcommand{\p}{\phi}
\newcommand{\pp}{\tilde \phi}
\newcommand {\pb}{\bar \phi}

\begin{document}
\title{Metastable de Sitter vacua from critical scalar theory}
\author{F. Loran}
\email{loran@cc.iut.ac.ir}
\author{E. Bavarsad}
 \email{bavarsad@ph.iut.ac.ir}
 \affiliation{Department of  Physics, Isfahan University of Technology
 (IUT), Isfahan,  Iran}

\begin{abstract}
 Studying the critical scalar theory in four dimensional Euclidean
 space with the potential term $-g\phi^4$ we show that the theory
 can not be analytically continued through $g=0$ from $g<0$ region to $g>0$
 region. For $g>0$ although energy is not bounded from below but there exist a
 classical trajectory with an $\mbox{AdS}_5$ moduli space,
 corresponding to a metastable local minima of the action.
 The fluctuation around this solution is governed by a minimally coupled scalar theory
 on four dimensional de~Sitter background with a reversed Mexican hat potential.
 Since in the weak coupling limit, the partition function
 picks up contribution only around classical solutions, one can assume
 that our de~Sitter universe corresponds to that local minima which lifetime increases
 exponentially as the coupling constant tends to zero. Similar results is obtained in
 the case of critical scalar theory coupled to U(1) gauge field which is essential
 for people living on flat Euclidean space to observe a de Sitter background by optical
 instruments.
\end{abstract}

\pacs{11.25.Hf, 98.80.-k,04.62.+v,11.25.Db} \maketitle

\section{Introduction}
 Recently, we showed that the fluctuations  of the scalar field around the
 classical trajectory of massless $\p^4$ model in four dimensional flat
 Euclidean spacetime is governed by a conformally
 coupled scalar field theory in four dimensional de Sitter background \cite{Solitons}.
 This result is interesting due to its uniqueness. In four dimensions, in principle,
 one can consider two classes of critical (classically scale-free ) scalar field
 theories i.e. massless $\phi^4$ models on Euclidean (Minkowski) spacetime
 with $g$, the coupling constant, either positive or negative (we assume the potential
 $V(\p)=-\frac{g}{4}\p^4$). Although scalar theory with $g>0$ seems
 to be not physical as energy is not bounded from below but as is
 shown in \cite{Solitons} in this case, the Euler-Lagrange equation of motion of
 the scalar theory on Euclidean space has an interesting classical solution say $\p_0$
 with finite action
 $S[\p_0]\sim g^{-1}$. Interestingly, as far as we are considering real scalar field
 theories the model with the physical potential i.e. the case of  $g<0$ has not such a
 solution, see Eq.(\ref{soliton}). We should clarify that from our previous viewpoint
 \cite{Solitons}, for $g<0$, one can still consider a solution like
 $\p_0$ obtained by an analytic continuation from $g>0$ to
 $g<0$ region. But such a solution is singular on the surface of a
 sphere which radius is proportional to $g$.
 Consequently the action $S[\p_0]$ is infinite and $\p_0$ can not be considered as a
 classical trajectory. If one ignores this problem and follows the
 calculations one obtains a conformally coupled scalar
 field on $\mbox{AdS}_4$ background. The reason why we do not follow our previous
 point of view turns back to our new machinery for constructing $\p_0$ from the first
 principles explained in section \ref{Cr}. Of course similar singularities
 appear even for $g>0$, when one switches to Minkowski spacetime by Wick
 rotation $x^0\to ix^0$. But in this case ($g>0$) one can argue that the singularity
 is beyond the horizon of observers living in the corresponding de~Sitter
 spacetime and consequently is safe. We do not study the case of Minkowski spacetime
 in this paper.

 As is shown in \cite{Solitons} the information geometry of the moduli of $\p_0$
 is  $\mbox{AdS}_5$ ($\mbox{dS}_5$) if $g>0$ ($g<0$) which resembles
 the information geometry of  $k=1$ $SU(2)$ instantons \cite{Blau}
  \footnote{$V(\p_0)$ can be shown to be proportional
 to the SU(2) one-instanton density \cite{Blau}.}. The moduli
 here are the location of the center of $\p_0$ and its size and
 are consequences of the invariance of the action under rescaling and
 translation. The similarity between this
 solution and SU(2) instantons can be explained in terms of the 't Hooft $\phi^4$ ansatz for instantons
\cite{Thooft}.

  In this paper focusing on the case $g>0$ we show that $\p_0$ can be responsible for
  viewing a metastable de Sitter background. We first show
  that $\p_0$ is a metastable local minima of the action.
  Then we show that the weak coupling limit $g\to{^+0}$ is
  equivalent to the classical ($\hbar\to0$) limit therefore in that
  limit the partition function picks up contribution only around the
  classical trajectory $\p_0$. As we said before, fluctuations around
  $\p_0$ is governed by a scalar theory on a de Sitter background with a
  reversed Mexican potential. The hight and width of the barrier is
  proportional to $R$ and $g^{-1}$. Here $R$ is the scalar curvature
  of the universe which can not be determined in our model but
  one can show that the lifetime of this local
  minima is proportional to $e^{g^{-1}}$.

  Summarizing these results we verify that the critical scalar theory can not be
  analytically continued (at least through $g=0$) from $g<0$ region to $g>0$ region.
  These results are strictly important from perturbation theory point of view.
  In the $g\to{^-0}$ limit, valid perturbations are around $\phi=0$ in a {\em flat } Euclidean
  background but in the $g\to{^+0}$ limit perturbations are around $\phi=0$ in an
  Euclidean {\em de~Sitter }  background. More explicitly at $g={0^-}$,
  the theory is simply a free massless scalar theory on flat Euclidean space
  but at $g={0^+}$ due to $\hbar$ considerations the theory is a (free and
  stable)  conformally coupled  scalar theory on an Euclidean de Sitter
  background. See sections \ref{Cr} and \ref{dS}.

  But why we are interested in de Sitter background. In fact WMAP results
  \cite{WMAP} combined with earlier cosmological observations
 shows that we are living in an accelerating universe. Constructing
 four dimensional de Sitter vacuum as a string theory (M-theory) solution
 has been a long standing challenge. In \cite{KKLT} KKLT constructed
 a  metastable de Sitter vacua of type IIB string theory by adding
 $\overline{\mbox{D3}}$-branes  to the GKP \cite{GKP} model of
 highly warped IIB compactifications with nontrivial NS and
 RR three-form fluxes. The mean lifetime of KKLT solution is $10^{10^{120}}$ years.
 KKLMMT constructed a model of inflation by adding mobile D3 branes to the KKLT
 solution \cite{KKLMMT}. In such models the inflaton (the position of D3 brane deep inside the warped
 throat geometry) is a conformally coupled scalar in the effective
 four-dimensional geometry. Therefore the mass of such a scalar $m^2_\phi$ is close to
 $2H^2=\xi R$ where $H^2$ is the Hubble parameter, $\xi=\frac{1}{6}$
 (in four dimension) is the conformal coupling constant and $R$ is the curvature
 of de Sitter space. As is uncovered in \cite{KKLMMT} this does not meet
 the observational requirement $m^2_\phi\sim 10^{-2}H^2$ \footnote{In D3/D7 model of
 inflation one does not encounter the $m^2_\phi\sim H^2$ problem.
 But here in contrast to the model of D3$/\overline{\mbox{D3}}$ inflation in the
 highly warped throat, there is no natural mechanism for suppressing
 the contribution of cosmic strings formed at the end of
 inflation to the CMB anisotropy. See \cite{DHKLZ} for a solution to
 this problem.}. Similar considerations show that our model can not
 be a successful model for inflation since in the $g\to{^+0}$ limit we also
 obtain a conformally coupled scalar theory on de Sitter background.

 Another problem in our model is the existence of a continuum of
 de Sitter bubbles, given by the location of the center of $\p_0$ and
 its size. Naively this number is proportional to the volume of the
 $\mbox{AdS}_5$ moduli space.
 If $\p_0$ is responsible to observe, say, by {\em optical} methods
 a de~Sitter geometry (see section \ref{U1}), a natural question is to ask which
 bubble we live in. Studying the variation of action
 around the $\p_0$ solution, we have verified, by numerical calculations that smaller bubbles
 are more stable than larger ones. Consequently there is a
 transition: larger bubbles decay to smaller ones and probably finally there remains a
 gas of zero size bubbles. The mechanism of such a
 decays  is not clear to us yet but its phenomenology, might be similar
 to that of the discretuum of possible de Sitter vacua in KKLT models
 \cite{fall}.

 As we said before our theory can not do any prediction about the size of our
 de Sitter universe or the nature of the scalar field but it predicts that
 the lifetime of this metastable vacua is $\tau\sim e^{g^{-1}}$.
 The model is interesting due to its uniqueness and symmetries which brings hopes to
 be constructible from a fundamental theory like string theory.

 The organization of the paper is as follows.
 In section \ref{Cr}, we study the critical scalar theory in $D=4$ Euclidean
 space and $\p_0$, the exact solution to
 the corresponding Euler-Lagrange equation of motion.  Considering the
 weak coupling limit we show that the partition function only
 picks~up contribution around $\p_0$. In section \ref{dS}, we show that the
 critical scalar theory considered as the action for (not essentially small)
 fluctuations around $\p_0$ is a scalar theory on $D=4$ de Sitter
 background. Consequently we verify that in the weak coupling limit
 the critical scalar theory of section \ref{Cr}is in fact a metastable scalar theory
 on de Sitter background with finite lifetime increasing exponentially as the
 coupling constant decreases. In section \ref{U1} we generalize our model to scalar theory
 coupled to U(1) gauge field. Such a generalization is essential as it shows how
 by optical observations people living in a flat Euclidean space can view
 a de~Sitter geometry for their universe.
 \section{Critical scalar theory in $D=4$ Euclidean
 space}\label{Cr}
 In this section we study scalar field theories in four dimensional Euclidean space
 invariant under rescaling transformation. By rescaling we mean a coordinate transformation
 $x\to x'=\lambda x$, $\lambda>0$. Requiring the kinetic term of a scalar theory to
 be invariant under rescaling, one verifies that the scalar field should be transformed as
 $\p(x)\to\p'(x')=\lambda^{-1}\p(x)$. A scale free theory by
 definition is a theory given by an action $S$ invariant under
 rescaling. In addition to the Kinetic term which variation is a total derivative,
 we search for polynomials $V(\p)$ in $\p$ as the potential term such that $\delta V=0$,
 up to total derivatives. Such polynomials exist only in three, four and six
 dimensions. In the case of our interest i.e. $D=4$,
 $V(\p)=\frac{-g}{4}\p^4$. Here $g$, the coupling constant is some arbitrary real constant
 which is by construction invariant under rescaling. Such a scalar model is called
 critical as it is scale-free and its correlation length is
 infinite. The corresponding Euler-Lagrange equation of motion is a
 nonlinear Laplace equation $\nabla^2 \p+g\p^3=0$. One can easily
 show that for $g>0$, a solution of the non-linear Laplace equation
 is \cite{Solitons,Fubini},
 \be
 \p_0(x;\beta,a^\mu)=\sqrt{\frac{8}{g}}
 \frac{\beta}{\beta^2+(x-a)^2},
 \label{soliton}
 \ee
 where $(x-a)^2=\delta_{\mu\nu} (x-a)^\mu(x-a)^\nu.$ $\beta$ and $a^\mu$ are
 undetermined parameters describing the
 the size and location of $\p_0$. These moduli are consequences
 of symmetries of the action i.e. invariance under rescaling and
 translation. The information geometry of the moduli space, given by
 Hitchin formula \cite{Hit}
 \be
 {\cal G}_{IJ}=\frac{1}{N}\int d^4x
 {\cal{L}}_0\partial_I\left(\log{\cal{L}}_0\right)
 \partial_J\left(\log{\cal{L}}_0\right),
 \label{InG}
 \ee
 is an Euclidean $\mbox{AdS}_5$ space \cite{Solitons},
 \be
 {\cal G}_{IJ}d\theta^Id\theta^J=\frac{1}{\beta^2}\left(d\beta^2+d
 a^2\right).
 \label{Mod-met}
 \ee
 $I=1,\cdots,5$ counts space directions of moduli
 space $\theta^I\in\{\beta,a^\mu\}$, $N=\frac{4^3}{5}\int d^4 x {\cal L}_0$ is
 a normalization constant and ${\cal{L}}_0=\frac{g}{4}\p_0^4$ is the Lagrangian density
 calculated at $\p=\p_0$.
 The moduli $a^\mu$ are present since the action is invariant under
 transformation. The existence of $\beta$ is the result of invariance under rescaling.
 To see this let us consider scale-free fields i.e. those fields that satisfy the
 relation $\delta_\epsilon\p=0$. Here $\delta_\epsilon\p(x)=\p'(x)-\p(x)$ is the
 infinitesimal scale transformation given by $\lambda=1+\epsilon$ for some
 infinitesimal $\epsilon$. To this aim we first note that rescaling leaves the
 origin ($x=0$) invariant. Consequently $\p(0)$ is distinguished from the values of
 the field at the other points since $\p(0)\to \p'(0)=\lambda^{-1}\p(0)$. Therefore,
 it is plausible to make the dependence of scalar fields on their values at the origin
 explicit and represent the rescaling transformation by
 $\p(x;\p(0))\to\p'(x;\p(0))=\lambda^{-1}\p(\lambda^{-1}x;\lambda\p(0))$.
 Defining $\beta^{-1}=\p(0)$, one can show that
 $\delta_\epsilon\p(x)=-\epsilon(1+x^I\partial_I)\p(x)$, where
 $x^I\in\{x^\mu,\beta\}.$ The $SO(4)$ invariant solutions of
 equation $\delta_\epsilon\p=0$ satisfying the condition
 $\p(0;\p(0))=\p(0)$ are $\p_k=c \beta^{-1}\left(\frac{\beta}{\sqrt{\beta^2+x^2}}
 \right)^{k+2}$ where $c$ is some arbitrary constant. It is easy
 to see that for $c=\sqrt{\frac{8}{g}}$, $\p_0$, among the others, is
 the solution of classical equation of motion.

 Now it is time to show that $\p_0$ is a metastable local minima of
 the action. Since $\p_0$ is a solution of Euler-Lagrange equation
 of motion $\delta S=0$ it is a local  extremum of the action\footnote{Of course
 from $\delta S=0$ we can only conclude that $\p_0$ is a stationary point
 and not necessarily a local extremum. We continue by assuming that $\p_0$ is a
 local extremum. This assumption can be proved following the results of
 section \ref{dS}.}. So it is enough to show that there are field variations
 $\p_0\to \p_\eta=\p_0+\epsilon
 \eta$ for ${\cal C}^1$ functions $\eta$ vanishing as $x\to\infty$
 such that $\delta S=c_\eta\epsilon^2+{\cal O}(\epsilon^3)$ for some
 real positive constant $c_\eta$. For simplicity one can assume
 $\eta=\left(\frac{1}{1+x^2}\right)^n$, $g=8$, $b=1$ and calculate
 $\delta S=S[\phi_\eta]-S[\p_0]$ for some integers $n$. One recognizes that $c_n>0$ for
 $n>5$, though it is negative for $0<n\le 5$. A good sign for metastability
 of the action at $\p_0$. Section \ref{dS} provides an exact proof for this claim.
 Another interesting observation is that bubbles with
 larger size are less stable than those with smaller size. This
 can be checked noting that the size of a bubble is proportional to
 $\beta^{-1}$. By repeating the above calculations one easily verifies
 that for example for $b=3$ $c_n>0$ even for $n=3$.
 Unfortunately without analytic data all these observations make only a
 rough picture of the phenomena which can not be used to make
 an exact statement.
 \subsection*{The weak coupling limit}
 To study the weak coupling limit of the theory one can scale
 $g\to \frac{g}{\lambda^2}$ and consider the $\lambda\to\infty$
 limit, while keeping fields $\p$ undistorted. To this aim we do the
 following transformations,
 \be
 \begin{array}{l}
 x\to\lambda x,\\
 \p\to\lambda^{-1}\p,\\
 g\to g,\\
  \end{array}
 \hspace{1cm}\mbox{and}\hspace{1cm}
 \begin{array}{l}
 x\to x,\\
 \p\to\lambda\p,\\
 g\to \lambda^{-2}g,\\
 \end{array}
 \ee
 which results in the desired transformation:
 \be
 \begin{array}{l}
 x\to \lambda x,\\
 \p\to\p,\\
 g\to \lambda^{-2}g,\\
 S\to \lambda^2 S,
 \end{array}
 \ee
 More explicitly due to invariance under rescaling we have,
 \be
 \int D[\p]\ e^{-\frac{1}{\hbar}S[\phi;\lambda^{-2}g]}=\int D[\p]\
 e^{-\frac{\lambda^{2}}{\hbar}S[\phi,g]}.
 \ee
 To calculate the partition function one can instead of the
 transformation $S\to \lambda^2 S$ assume that $S\to S$ but
 $\hbar\to \lambda^{-2}\hbar$.

 Consequently the $g\to{^+0}$
 limit is equivalent to $\hbar\to {^+0}$ limit \footnote{Of course we are also blowing
 our universe as $x\to \lambda x$. Since the flat Euclidean space we considered is not compact
 it does not seems to cause any problem at this level. In the case $n$-point
 functions one should note that non-coincident points go to infinity with respect
 to each other}. Therefore in the weak
 coupling limit the partition function picks up contribution only
 from trajectories close to $\p_0$. This key observation when we
 study coupling to $U(1)$ gauge field proves why people living in a flat
 Euclidean space with a critical scalar field at $g=0^+$ observe a
 de~Sitter universe.
 \section{$\p_0$ as a de Sitter background}\label{dS}
 In this section we show that fluctuations around $\p_0$ are governed by a scalar
 theory on de~Sitter background. This section is a review
 of the calculations made in \cite{Solitons}.
 In order to study the fluctuations around $\p_0$ one should rewrite the action
 \be
 S[\p]=\int d^4x
 \left(\frac{1}{2}\delta^{\mu\nu}\partial_\mu\p\partial_\nu\p-
 \frac{g}{4}\p^4\right),
 \label{action}
 \ee
  in terms of new fields $\pp=\p-\p_0$. In this way one obtains a new
  action,
 \be
 S[\p]=S[\p_0]+S_{\mbox{free}}[\pp]+S_{\mbox{int}}[\pp],
 \label{act}
 \ee
 where $S[\p_0]=\int d^4x {\cal L}_0=\frac{8\pi^2}{3g}$, and
 \be
 S_{\mbox{free}}[\pp]=\int d^4 x \left(
 \frac{1}{2}\delta^{\mu\nu}\partial_\mu\pp\partial_\nu\pp+\frac{1}{2}M^2(x)\pp^2\right),
 \label{free}
 \ee
 in which,
 \be
 M^2(x)=-3g \p_0^2
 =-24\frac{\beta^2}{\left(\beta^2+(x-a)^2\right)^2}.
 \label{mass}
 \ee
 The mass dependent term can be interpreted as interaction with $\p_0$ background.
 Now recall that in general, by inserting $\pp=\sqrt{\Omega}\pb$ and
 $\delta_{\mu\nu}=\Omega^{-1}g_{\mu\nu}$ in the action
 $S[\pp]=\int d^4x \frac{1}{2}\delta^{\mu\nu}\partial_\mu\pp\partial_\nu\pp$,
 one obtains,
 \be
 S[\pp]=\int d^4x \sqrt{g}\left(\frac{1}{2}g^{\mu\nu}\partial_\mu\pb\partial_\nu\pb+
 \frac{1}{2}\xi R\pb^2
 \right),
 \ee
 i.e. a scalar theory on conformally flat background given by the metric
 $g_{\mu\nu}$ in which  $\Omega>0$ is an arbitrary
 ${\cal C}^\infty$ function. $R$ is the scalar curvature of the
 background and $\xi=\frac{1}{6}$ is the conformal coupling
 constant. For details see \cite{Ted} or appendix C of
 \cite{Solitons}.
 Thus, defining $\pb=\Omega^{\frac{-1}{2}}\pp$,
 one can show that $S_{\mbox{free}}[\pp]$ given in Eq.(\ref{free})
 is the action of the scalar
 field $\pb$ on some conformally flat background
 with metric $g_{\mu\nu}=\Omega\delta_{\mu\nu}$:
 \be
 S_{\mbox{free}}[\pp]=\int d^4 x \sqrt{\left|g\right|}
 \left(\frac{1}{2}g^{\mu\nu}\partial_\mu\pb
 \partial_\nu\pb+\frac{1}{2}(\xi R+m^2)\pb^2\right).
 \label{Curvedaction}
 \ee
 Here, $m^2\Omega=M^2(x)$, where $m^2$ is the mass of $\pb$
 (undetermined) and $M^2(x)$ is given in Eq.(\ref{mass}).
 This result is surprising as one can show that the Ricci tensor
 $R_{\mu\nu}=\Lambda g_{\mu\nu}$, where $\Lambda=-\frac{m^2}{2}>0$ as far as $\Omega>0$.
 Consequently $\pb$ lives in a four dimensional
  de~Sitter space which scalar curvature $R=-2m^2$.
  The interacting part of the action,
  $S_{\mbox{int}}[\pp]=\int d^4x\sqrt{\left|g_{\mu\nu}\right|}{\cal L}_{\mbox{int}}$
  is well-defined in terms of $\pb$ on the corresponding $\mbox{dS}_4$:
  \be
  {\cal L}_{\mbox{int}}=
  -g\sqrt{\frac{-m^2}{3g}}\pb^3-\frac{g}{4}\pb^4.
  \ee
  Interestingly after a shift  of the scalar field $\pb\to\pb-\sqrt{\frac{-m^2}{3g}}$
  the action  (\ref{act}) can be written in the $\mbox{dS}_4$ as follows:
  \be
  S[\pb]=\int d^D x \sqrt{\left|g\right|}
 \left(\frac{1}{2}g^{\mu\nu}\partial_\mu\pb
 \partial_\nu\pb+\frac{1}{2}(\xi
 R)\pb^2-\frac{g}{4}\pb^4\right).
  \label{R}
  \ee
  This is a scalar theory in a de Sitter background with reversed
  Mexican hat potential.   $\pb=0$ corresponds to the local minima of
  the potential which distance to the center of the hill (the
  location of $\p_0$) is $\sqrt{\frac{\xi R}{g}}$. The hight of
  the hill is $\frac{\xi^2 R^2}{4g}$. The lifetime of the
  metastable vacua can be estimated using the WKB method:
  the transition rate $\Gamma$ is
 \be
  \log \Gamma\sim -\Delta S,\hspace{1cm}\Delta S\sim V_4\frac{(\xi R)^2}{g}
  \ee
  in which $V_4\sim R^{-2}$ is the volume of the ${\cal S}^4$, the Euclidean de Sitter space.
  Consequently the lifetime $\tau\sim e^{g^{-1}}$.
 \section{Critical scalar theory coupled to $U(1)$ gauge
 field}\label{U1}
 In this section we study complex critical scalar theory on
 Euclidean space coupled to $U(1)$ gauge field $A_\mu$,
 \be
 S=\int d^4
 x\left(\left|D_\mu\p\right|^2-\frac{g}{2}\left|\p\right|^4\right)+S_A,
 \label{Uact}
 \ee
 where $D_\mu=\partial_\mu+i e A_\mu$ is the covariant derivative
 and
 $S_A=-\frac{1}{4}F_{\mu\nu}F_{\rho\sigma}\delta^{\rho\mu}\delta^{\sigma\nu}$.
 It is easy to verify that $A_\mu=0$ and $\p=\p_0$ is a solution of
 the Euler-Lagrange equation of motion. Inserting $\pp=\p-\p_0$ in
 the Eq.(\ref{Uact}), one obtains
 \be
 S=S[\p_0]+{\tilde S}[\pp]+S_{int}+S_A,
 \ee
 where
 \bea
 S_{int}&=&\int d^4x \delta^{\mu\nu}\left[\left(ie A_\mu(\pp+\p_0)\partial_\nu({\pp}^*+\p_0)
 +c.c.\right)\right.\nn\\
 &&\ \ \ \ \ \ \ \ \ \ \ \ \ \ +\left. e^2 A_\mu
 A_\nu\left|\pp+\p_0\right|^2\right],\nonumber\\
 {\tilde S}[\pp]&=&\int d^4 x
 \left(\frac{1}{2}\left|\partial_\mu(\pp+\p_0)\right|^2-
 \frac{g}{4}\left|\pp+\p_0\right|^4\right).
 \eea
 Inserting $\pb=\Omega^{-\frac{1}{2}}\pp$ where $\Omega$ is defined as before by the
 relation $\p_0=\sqrt{\frac{-m^2}{3g}\Omega}$ one obtains,
 \bea
 S_{int}&=&\int d^4 x \frac{1}{2}\Omega\delta^{\mu\nu}\left[ie A_\mu\left(\pb+\sqrt
 {\frac{-m^2}{3g}}\right)\partial_\nu{\pb}^*
 +c.c.\right.\nn\\&& \ \ \ \ \ \ \ \ \ \ \ \ \ \ \left.
  +e^2 A_\mu A_\nu\left|\pb+\sqrt
 {\frac{-m^2}{3g}}\right|^2\right],
 \eea
 and
 \bea
 {\tilde S}[\pp]&=&\int d^4 x
 \left(\frac{1}{2}\Omega\left|\partial_\mu\pb\right|^2
 -\frac{g}{4}\left|\pb+\sqrt
 {\frac{-m^2}{3g}}\right|^4\right.\nn\\&&\ \ \ \ \ \ \ \ \ \ \left.
 -\frac{1}{2}\sqrt{\Omega}\nabla^2\sqrt{\Omega}\left|\pb+\sqrt
 {\frac{-m^2}{3g}}\right|^2\right),
 \eea
 where $\nabla^2=\delta^{\mu\nu}\partial_\mu\partial_\nu$. Defining
 $g_{\mu\nu}=\Omega\delta_{\mu\nu}$ and noting that
 $-\sqrt{\Omega}\nabla^2\sqrt{\Omega}=\sqrt{g}\xi R$, after an
 obvious shift $\pb\to\pb-\sqrt {\frac{-m^2}{3g}}$, one obtains,
 \bea
 S&=&\int d^4 x\sqrt{g}\left(\frac{1}{2}g^{\mu\nu}D_\mu\pb
 D_\nu{\pb}^*+\frac{1}{2}\xi
 R\left|\pb\right|^2-\frac{g}{4}\left|\pb\right|^4\right)\nn\\&+&S_A,
 \eea
 It is known that in $D=4$, $S_A$ is invariant under conformal
 transformation $\delta_{\mu\nu}\to \Omega\delta_{\mu\nu}$ and $A_\mu\to A_\mu$
 thus one can write the action $S_A$ equivalently as follows,
 \be
 S_A=-\frac{1}{4}\int d^4 x
 \sqrt{g}g^{\mu\rho}g^{\nu\sigma}F_{\mu\nu}F_{\rho\sigma},
 \ee
 where $F_{\mu\nu}=\partial_\mu A_\nu-\partial_\nu A_\mu$ is the
 field strength in the de Sitter background.
 Consequently at $g=0^+$ the theory given by the action (\ref{Uact})
 is a conformally coupled scalar theory minimally coupled to $U(1)$
 gauge field on de~Sitter background. Therefore at $g=0^+$ using optical
 instruments people living on flat Euclidean space observe an accelerating universe.
  \section*{Acknowledgement}
 The financial support of Isfahan University of Technology (IUT) is
 acknowledged.
 
\end{document}